\documentclass{emulateapj}
\usepackage{verbatim}

\shorttitle{MAGIC observation of Willman~1}

\begin{document}

\title{UPPER LIMITS ON THE VHE GAMMA-RAY EMISSION FROM THE WILLMAN 1
  SATELLITE GALAXY WITH
THE MAGIC TELESCOPE}

%
\author{
E.~Aliu\altaffilmark{a},
H.~Anderhub\altaffilmark{b},
L.~A.~Antonelli\altaffilmark{c},
P.~Antoranz\altaffilmark{d},
M.~Backes\altaffilmark{e},
C.~Baixeras\altaffilmark{f},
S.~Balestra\altaffilmark{d},
J.~A.~Barrio\altaffilmark{d},
H.~Bartko\altaffilmark{g},
D.~Bastieri\altaffilmark{h}
% ...
}
\altaffiltext{a} {IFAE, Edifici Cn., Campus UAB, E-08193 Bellaterra, Spain}
\altaffiltext{b} {ETH Zurich, CH-8093 Switzerland}
\altaffiltext{c} {INAF National Institute for Astrophysics, I-00136 Rome, Italy}
\altaffiltext{d} {Universidad Complutense, E-28040 Madrid, Spain}
\altaffiltext{e} {Technische Universit\"at Dortmund, D-44221 Dortmund, Germany}
\altaffiltext{f} {Universitat Aut\`onoma de Barcelona, E-08193 Bellaterra, Spain}
\altaffiltext{g} {Max-Planck-Institut f\"ur Physik, D-80805 M\"unchen, Germany}
\altaffiltext{h} {Universit\`a di Padova and INFN, I-35131 Padova, Italy}
% ...

\begin{abstract}
We present the result of the observation of the ultrafaint dwarf
galaxy Willman 1 performed with the 17 m MAGIC telescope during 15.5
hr between March and May 2008. No significant $\gamma$-ray emission was
found.We derived upper limits of the order of $10^{-12}$ ph cm$^{-2}$ s$^{-1}$ on
the integral flux above 100 GeV, which we compare with
predictions from several of the established neutralino benchmark
models in the mSUGRA parameter space. The neutralino annihilation
spectra are defined after including the recently quantified
contribution of internal bremsstrahlung from the virtual sparticles
that mediate the annihilation. Flux boost factors of three orders of
magnitude are required even in the most optimistic scenario to match
our upper limits. However, uncertainties in the dark matter intrinsic
and extrinsic properties (e.g., presence of substructures, Sommerfeld
effect) may significantly reduce this gap. 
\end{abstract}


\section{Introduction}
Dwarf spheroidal galaxies (dSphs) are believed to be the smallest
(size $1\sim$~kpc), faintest (luminosities $10^2-10^8$~L$_\odot$)
astronomical objects whose dynamics are dominated by dark matter
\citep[DM;][and references therein]{Gilmore:2008}.
They are found as satellites orbiting in the gravitational
field of a larger host galaxy (e.g., the Milky Way (MW)).
Their member stars show large circular velocities and velocity
dispersions that, combined with their modest spatial extent, can be
interpreted with the presence of a large DM halo of the order of
$10^5-10^8$~M$_\odot$. In recent years, the Sloan Digital Sky Survey
\citep[SDSS;][]{York:2000} led to the discovery of a new population of
MW satellites, comprising about as many new objects as were
previously known
\citep{Belokurov:2004,Willman:2005a,Zucker:2006,Irwin:2007,Walsh:2007}.
This population of 
extremely low-luminosity galaxies is very interesting for DM searches
and to study the galaxy formation at the lowest mass scale. The
existence of a new class of ultrafaint MW satellites is also relevant
because it provides a partial solution for the so-called  missing
satellite problem \citep{Klypin:1999, Simon:2007, Strigari:2007,
  Madau:2008} by partially filling the gap between the 
predicted and the measured number of galactic subhalos. 
In this family is Willman~1, discovered by \citet{Willman:2005a}
and soon established as potentially the most DM-dominated dSph
satellite of the MW \citep{Martin:2007,SanchezSalcedo:2007, Strigari:2008}. 

The physics of DM has gathered much interest in recent years, in
particular after \emph{Wilkinson Microwave Anisotropy Probe} \citep{Spergel:2007, Komatsu:2008}
measured its relic density with great precision. It is generally
believed that DM manifests itself as a general class of
weakly interacting massive particles that includes several candidates
which satisfy both experimental constraints and theoretical
requirements~\citep[see][and references therein]{Bertone:2005a}. Among
them, one of the best theoretically motivated, for whom the 
relic density is calculated without fine tuning from its nature, is
the $neutralino$, arising in SuperSymmetric (SUSY) theories 
beyond the Standard Model~\citep{Wess:1974,Haber:1985} and
in particular in the mSUGRA extension~\citep{Chamseddine:1982}.
The mSUGRA neutralino annihilations can be observed through the
production of $\gamma$-rays. The main emission comes from secondary
products of hadronization processes and from final state radiation. In
addition, line emissions are found through direct processes such as
$\chi\chi\rightarrow\gamma\gamma$, and  $\chi\chi \rightarrow Z^0
\gamma$, which provide $\gamma$-rays of energies $E=m_\chi$ and
$E=m_\chi-m^2_{Z^0}/m^2_\chi$ respectively, even if those processes
are at the loop level and therefore strongly suppressed.

Recently~\citet{Bringmann:2008a}, following an earlier idea
from~\citet{Bergstrom:1989}, showed that in some regions of the mSUGRA  
parameter space, a hitherto neglected contribution to $\gamma$-ray emission
comes directly from charged sparticles mediating the annihilation into
leptons, in processes like $\chi\chi\rightarrow 
l^+l^-\gamma$. They defined this intermediate state radiation 
as \emph{internal bremsstrahlung} (IB). The IB mechanism
permits to restore the helicity balance in processes that would
otherwise be strongly forbidden, and supplies $\gamma$-rays toward
high energies ($E>0.6\;m_\chi$) of up to several orders of magnitude.
This boosted emission is particularly interesting for
ground-based $\gamma$-ray observation, as performed by Imaging Atmospheric
Cherenkov Telescopes (IACTs), which are sensitive above
$\sim100$~GeV where normally the IB boost takes place. 

As the $\gamma$-ray flux is proportional to the square of the DM density, IACTs 
focus on concentrated DM objects. 
Dwarf galaxy satellites of the MW represent very good candidates,
since they are the most DM dominated systems known 
in the universe, with very high mass-to-light ratios (M/L), close
distance and reduced $\gamma$-ray background from unresolved
conventional Galactic sources (i.e. stellar evolutionary remnants). 
Some dSphs have already been studied in $\gamma$-rays: Draco by
MAGIC~\citep{Albert:2008a} and  Whipple~\citep{Wood:2008}; UMi by
Whipple~\citep{Wood:2008} and  Sagittarius by
HESS~\citep{Aharonian:2007} without any significant observation of DM
annihilations and  
only flux upper limits were estimated.

In this paper, we report results of the observation of the sky region
around Willman~1 performed by the MAGIC telescope for a total of
$15.5$~hr between March and May 2008. After a brief description of
Willman~1 in Section~\ref{sec:willman}, in Section~\ref{sec:flux}, we estimate the
flux using benchmark models for the neutralino and a typical DM
density profile. In Section~\ref{sec:data}, we briefly
describe the MAGIC  telescope and the Willman~1 data sample. In
Section~\ref{sec:results},  we present and discuss the results of the
observation and we set upper limits for the flux. In 
Section~\ref{sec:conclusions}, we report our conclusions.

\section{Willman~1}
\label{sec:willman}
In 2004, Willman et al. discovered a new MW companion SDSS J1049+5103
\citep[$10^h40^m22.3^s$,$51^\circ03'03.6''$;][]{Willman:2005a} as a
faint overdensity of red, resolved stars, which was 
observed again the next year~\citep{Willman:2005b} and named
Willman~1. At that moment, this object represented  
the 10th dSph of the MW and, the first one discovered in 10
years. Further observations performed with the Keck/DEIMOS telescope
confirmed the SDSS results~\citep{Martin:2007}, while a more recent
observation is reported by~\citet{Siegel:2008}. 

Willman~1 is located at a distance of $38\pm7$~kpc in the Ursa Major
constellation. It is characterized by a very low number of resolved
stars, the total luminosity being $\mathrm{L}=855$~L$_\odot$, and a very small
half-light radius of $r_{1/2}=21\pm7$~pc, almost 2 orders of
magnitude smaller than other known dSphs. The source was defined by previous
authors as an ``extreme'' dwarf galaxy, because some of its 
characteristics lie between those typical for a globular cluster (GC)
and those expected in an extremely faint dSph. The large spread in
metallicity of its stars favors the dSph interpretation rather than
that of a GC, which would contain stars of a similar age and
metallicity~\citep{Martin:2007}, even if this evidence was recently
put under discussion by~\citet{Siegel:2008}. Willman~1 is the least
massive satellite galaxy known to date, with a total mass
($\mathrm{M}\sim5\times10^5\mathrm{M}_{\odot}$) about an order of
magnitude smaller than those of the least massive satellite galaxies
previously known. However, the corresponding mass-to-light ratio, 
M/L$\sim500-700$~M$_\odot/$L$_\odot$, is one of the highest of all dSphs. 
This makes Willman~1 one of the most attractive dSph galaxies to look 
for DM at present~\citep[see, e.g.,][]{Bringmann:2008b}, its predicted DM 
annihilation flux being  at least a factor of 3 larger than 
the second best DM candidate, according to recent work~\citep{Strigari:2007}.

\section{Theoretical modeling of the gamma-ray emission from Willman~1}
\label{sec:flux}

The $\gamma$-ray flux originating from DM particle annihilations can
be factorized into a contribution called the \emph{astrophysical
  factor} $J(\Psi)$ related to the morphology of the emission
region and a contribution called the \emph{particle physics factor}
$\Phi^{PP}$ depending on the candidate particle characteristics:
\begin{equation}\label{eq:flux}
\Phi(>E_0) = J(\Psi) \cdot \Phi^{PP}(>E_0)\;,
\end{equation} 

\noindent
where $E_0$ is the energy threshold of the detector and $\Psi$ the
angle under which the observation is performed.

\subsection{Astrophysical Factor}
\label{subsec:astro}
At present, a concise and exact characterization of the DM density
profile of Willman~1 is a delicate task, since observational data are still
scarce.  This paper is based on the studies from
\citet{Strigari:2008}, who modeled the profile by using only 47 stars
after removing 
those with unclear kinematics. Furthermore, to avoid membership
problems, only the  observational data related to the inner half of
the galaxy were taken into account. It is important to note that a
null/insignificant tidal stripping was assumed in order to carry out
the modeling of the DM distribution as a system in dynamical
equilibrium, a fact which  is still under debate.
For example, \citet{Willman:2006} claim the existence of strong
tidal debris, the evolution of the dwarf being strongly affected by
tidal interactions with the MW still now, although DM constitutes 90\%
of its total mass. In the same line, \citet{Martin:2007}, following
deep observations in the $r$ band, infer that Willman~1 may 
probably be surrounded by tidal tails. The authors give as a
plausible scenario that the dwarf could have been significantly
tidally stripped but only in the past, when the object was more luminous and
massive. At that age, Willman~1 could have lost most of its
outskirts, only the innermost regions surviving intact. This picture
would then allow the two contradictory arguments to coexist, since at least a
correct modeling for the core of the dwarf may be possible assuming
this region to be in dynamical equilibrium at present. 

A well established model for the DM distribution, and the only one so
far applied to Willman~1 \citep{Strigari:2008}, is the
Navarro-Frenk-White (NFW) density
profile~\citep{Navarro:1997}:
\begin{equation}\label{eq:rho}
\rho_{\mbox{\footnotesize{NFW}}}(r)\;=\;\rho_s
\left(\frac{r}{r_s}\right)^{-1}\left(1+\frac{r}{r_s}\right)^{-2}\;, 
\end{equation}

\noindent
where $\rho_s$ and $r_s$ are a typical scale density and radius
respectively. The astrophysical factor can be written as: 
\begin{equation}\label{eq:jpsi}
J(\Psi_0)=\frac{1}{4\,\pi}\int_{V}d\Omega\;\int_{los} d\lambda \;
[\rho^2(r)* B_{\vartheta_r}(\theta)]\;, 
\end{equation}

\noindent
where $\Psi_0$ denoted the direction of the target. The first integral
is performed over the spatial extension of the source, the second is
performed over the line of sight variable~$\lambda$. The density is
convoluted with the Gaussian function 
$B_{\vartheta_r}(\theta)$ describing the telescope angular resolution
where $\theta=\Psi-\Psi_0$ is the angular distance with respect to 
the center of the object. 
We remark that the integration of Equation~\ref{eq:jpsi} involves
foreground (MW halo) and extragalactic background whose contributions
can be substantial~\citep{Elsaesser:2005}.  

At a distance of 38~kpc, the scale radius corresponds to 
an extension of $0.54^\circ$ in the sky, which is well inside the MAGIC
field of view ($\sim3.5^\circ$), but is rather extended compared with
the telescope angular resolution of $0.1^\circ$.  This
evidence is mitigated by the fact that the main emission
still comes from the very core of the source, due to the very steep
NFW profile at the center and the square density dependence. For this
reason, we performed an analysis adapted for slightly extended
sources~\citep{Sitarek:2008}.  
To compute the astrophysical factor for Willman~1, we substitute
$r_s=0.18$~kpc and $\rho_s=4\times 10^8$ M$_\odot/$kpc$^3$
taken from~\citet{Strigari:2008} into Equation~\ref{eq:rho}, and by computing 
Equation~\ref{eq:jpsi} we obtain $J(\Psi_0)\sim3.5\times10^{17}$~GeV$^2$/cm$^5$. 
This value is possibly the largest over the rest of dSphs.

\subsection{Particle Physics Factor}
\label{subsec:phipp}
\begin{figure}[hbt!]
\centering
\includegraphics[width=0.99\linewidth]{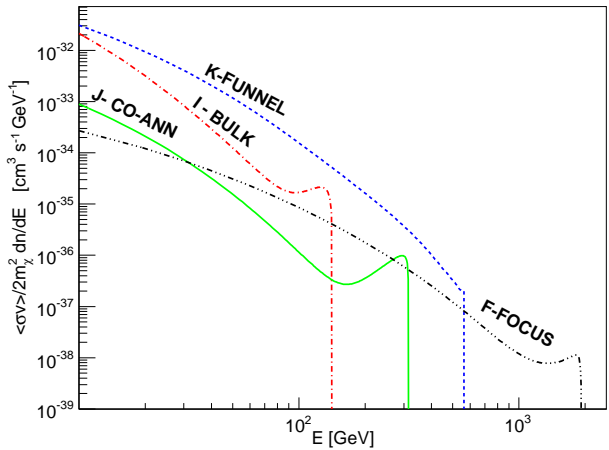}
\caption{Differential particle physics factor for the benchmarks
  models as in \citet{Bringmann:2008b}. Line gamma emissions are not
  included, since  their contribution to the flux is almost
  negligible. \label{fig:models}}  
\end{figure}

In many mSUGRA models, the lightest SUSY particle is
one of the four neutralinos ($\tilde\chi^0_{1\ldots4}$) which are linear
combinations of the Bino $\tilde{B^0}$, the Wino $\tilde{W^0}$ 
and the two neutral higgsinos $\tilde{H^0_u},\;\tilde{H^0_d}$. Only
five free parameters fully characterize mSUGRA: the 
scalar mass $m_0$ and the gaugino mass $m_{1/2}$ defined at the
unification scale, the trilinear scalar coupling $A_0$ and the ratio
$\tan\beta$ of the Higgs vacuum expectation values. In addition, one 
needs to define the sign of the Higgs mass parameter sign$(\mu)$. The
typical DM $\gamma$-ray annihilation spectrum is a continuum with
a sharp cutoff at the DM candidate mass and possibly with a bump at energies
larger than $0.6$~m$_\chi$ in case the IB is present (see
Figure~\ref{fig:models}). 

The particle physics factor can be written as a product of two terms,
the first depending only on the DM candidate mass and cross section,
and a second term, depending on the $\gamma$-ray spectra of annihilation, which
must be integrated above the energy threshold $E_0$ of 
the telescope:
\begin{equation}\label{eq:phipp}
\Phi^{PP}(>E_0) = 
\frac{\langle\sigma v_{\chi\chi}\rangle}{2\,m^2_\chi}
\, \int_{E_0}^{m_\chi} S(E)\;\textrm{d}E\;,
\end{equation}

\noindent
where $\langle\sigma v_{\chi\chi}\rangle$ is the total averaged
thermal cross section times the relative velocity of particles, 
$m_\chi$ is the DM particle mass, and the factor $2$ 
takes into account that the neutralino annihilates with itself. The
$\gamma$-ray an\-ni\-hi\-la\-tion spectrum is composed of different
contributions: $S(E)=\sum_i\,\textrm{d}N^i_\gamma/\textrm{d}E$ where
$\textrm{d}N^i_\gamma/\textrm{d}E$ is the spectrum of the $i$th
annihilation mode. 

The mSUGRA parameter space is conventionally described in a $m_0\oplus
m_{1/2}$ plane, after having fixed the other free parameters. Usually, four
zones are identified: the \emph{bulk region}, with low $m_0$
and $m_{1/2}$ and neutralino masses at around 100~GeV, the \emph{focus
  point} where $m_0$ and the neutralino are more massive, the \emph{funnel
  region} where both $m_0$ and $m_{1/2}$ take large values, and the 
\emph{co-annihilation tail} characterized by large $m_{1/2}$. A
neutralino shows different annihilation modes depending on the
location in this plane. A representative set of benchmarks was
defined by~\citet{Battaglia:2001,Battaglia:2004}. Hereafter, we use a subset of four  
slightly modified Battaglia models, as defined
by~\citet{Bringmann:2008b} using DarkSUSY 4.01~\citep{Gondolo:2004},
which include the contribution of IB in the 
computation of the  cross sections and spectra:
models {\it I', J', K', F*} for the bulk, co-annihilation, funnel and
focus point regions respectively. All the defining parameters, as well
as the derived $\Phi^{PP}(>E_0)$, are listed in Table~\ref{tab:bm}.

\begin{table*}[hbt!]
\begin{center}
\caption{\label{tab:bm}Definition of benchmark models as in
  \citet{Bringmann:2008b} and computation of the particle physics
  factor:  $m_{1/2}$ and $m_0$ [GeV] are the gaugino and scalar mass 
  respectively defined at the unification scale; $\tan\beta$ is the
  ratio of the Higgs expectation values;  $A_0$ [GeV] the trilinear coupling
  constant and $sign(\mu)$ the sign of the Higgs mass; $m_\chi$ [GeV] is the
  neutralino mass; $\langle\sigma v_{\chi\chi}\rangle$ [cm$^3$
  s$^{-1}$] is the cross section times the relative velocity of DM
  particles and $\Phi^{PP}(>100)$ [cm$^3$ GeV$^{-2}$ s$^{-1}$] is the  
  particle physics factor above 100~GeV.}
\vspace{2mm}
\begin{tabular}{ccccccccc}
\tableline\tableline
BM & $m_{1/2}$ & $m_0$ & $\tan\beta$ & $A_0$ & $sign(\mu)$ &  
  $m_\chi$ & $\langle\sigma v_{\chi\chi}\rangle$ &
  $\Phi^{PP}(>100)$ \\
\tableline
$I'$    & 350  & 181   & 35   & 0    & $+$ & 141  
& $3.62\times10^{-27}$ & $7.55\times10^{-34}$\\
$J'$    & 750  & 299   & 35   & 0    & $+$ & 316  
& $3.19\times10^{-28}$ & $1.23\times10^{-34}$\\
$K'$    & 1300 & 1001  & 46   & 0    & $-$ & 565  
& $2.59\times10^{-26}$ & $6.33\times10^{-33}$\\
$F^*$ & 7792 & 22100 & 24.1 & 17.7 & $+$ & 1926 
& $2.57\times10^{-27}$ & $5.98\times10^{-34}$\\
\tableline
\end{tabular}
\end{center}
\end{table*}

The total estimated flux due to DM annihilation computed according to
Equation~\ref{eq:flux} is given in Table~\ref{tab:results}. These estimations must
be taken with carefulness since there are some sources of uncertainty
that may largely affect the values for the predicted flux: (1) our
lack of knowledge of the DM density profile may change the
astrophysical factor by more than 1 order of magnitude; (2) the
presence of DM substructures in the dwarf may enhance the $\gamma$-ray flux
at least by a factor of $2-3$ according to $N-$body simulations
\citep{Diemand:2007a,Diemand:2007b, Kuhlen:2008} or even up to $20$
according to \citet{Martinez:2009}. Substructures are in fact
expected to be present not only in Willman 1 but also in any other DM
halo, since cold dark matter (CDM) halos are approximately self-similar until a cutoff 
scale mass which lies in the range of
10$^{-4}$-10$^{-12}$~M$_{\odot}$ \citep{Profumo:2006}; (3) the exclusion of the baryons in the modelization of the total
density profile. However, the 
effect of the adiabatic compression \citep{Prada:2004,Gnedin:2004},
although important for larger DM   
halos, will probably play a marginal role in the case of Willman 1,
given its relatively low amount of baryons even in the central
regions, where the effect is expected to be more important;
(4) the inclusion of the Sommerfeld effect, recently investigated by
\citet{Lattanzi:2008}. This effect predicts a general enhancement $S$ of
the flux, in case of very low DM particle velocities $v$, proportional
to the inverse of the particle velocity: $S\propto1/v$. In addition, for 
particular values of the DM mass and very small velocities, due to the
presence of bound states, the Sommerfeld effect would give rise to
resonances, which enhance the flux of a factor $S\propto 1/v^2$.  
Finally, we underline that our choice of
benchmarks does not 
scan the complete parameter space and different neutralinos could have
a larger expected flux.  A deeper study of the parameter space is
therefore very desirable.

\section{MAGIC data}
\label{sec:data}
The MAGIC (Major Atmospheric Gamma Imaging Cherenkov) telescope is
located on the Canary Island La Palma (2200 m asl, $28.45^\circ$N,
$17.54^\circ$W). MAGIC is currently the largest IACT, having a 17~m
diameter tessellated reflector dish. The faint Cherenkov light flashes
produced by atmospheric showers initiated by VHE $\gamma$-rays in the
top atmosphere are reflected into an ellipsoidal image in the focal
plane of the telescope, where a camera consisting of 577
photomultipliers (pixels) records the image.  More details can be found in
\citet{Cortina:2005}.  

MAGIC observed Willman~1 between March and May 2008. The source was
surveyed at zenith angles between $22^\circ$ and $30^\circ$, which
guarantees the lowest energy threshold. The source was tracked for
$16.8$~hr plus another  $9.3$~hr in OFF observation mode,
i.e. pointing to a dark patch in the sky close to Willman~1 where no
$\gamma$-ray emission is expected, for background estimation. The main
background Cherenkov telescopes have to deal with is produced by
cosmic hadronic particles impinging onto the top atmosphere and
generating electromagnetic subshowers that can mimic pure
$\gamma$-rays showers, and by the night sky background. Background
events are partly rejected at the trigger level and in the 
off-line analysis event selection, following a 
procedure called gamma/hadron ($g/h$) separation. 

The analysis proceeds as
follows \citep[for a detailed description, see][]{Albert:2008b}: data
are calibrated and the number of photoelectrons per pixel extracted
\citep{Albert:2008c}, then an 
image cleaning selects pixels with at least six photoelectrons (three
photoelectrons in the boundary of the image). Additional
suppression of pixels containing 
noise is achieved by requesting a narrow time coincidence between 
adjacent pixels ($\sim7$~ns). Based on the Hillas parameterization
algorithm \citep{Hillas:1985}, the shower parameters are
reconstructed. The hadronic background is suppressed
with a multivariate method, the Random Forest
\citep{Breiman:2001, Albert:2008d}, that uses the Hillas
parameters  
 to define an estimator called \emph{hadronness} by comparison
with Monte Carlo (MC) $\gamma$-ray simulations. The hadronness
expresses the likeness of an event to be a hadron and runs from 0
for gammas to 1 for hadrons. The Random Forest method is also 
used to estimate the energy of a reconstructed shower and
the energy threshold is defined by the peak of the distribution of
reconstructed MC gamma events. The $g/h$ separation is optimized
on a real data sample from the Crab Nebula, a supernova remnant and
one of the brightest and stable $\gamma$-ray emitters, which is taken as
standard candle in very high energy $\gamma$-ray astronomy. The
optimization yields a best set of cuts 
in the Hillas parameters which defines the gamma and hadron acceptance
of the analysis. In our case, the optimal set of cuts is obtained
for an energy threshold of 100~GeV and a hadronness cut of $0.15$. 
The overall data quality is very high, with only $7$\% data
rejection, resulting in 15.5~hours effective observation
time. Independent cross-checks were performed on the data giving
compatible results.

\section{Results and Discussion}
\label{sec:results}
\begin{figure}[hbt!]
\centering
\includegraphics[width=0.99\linewidth]{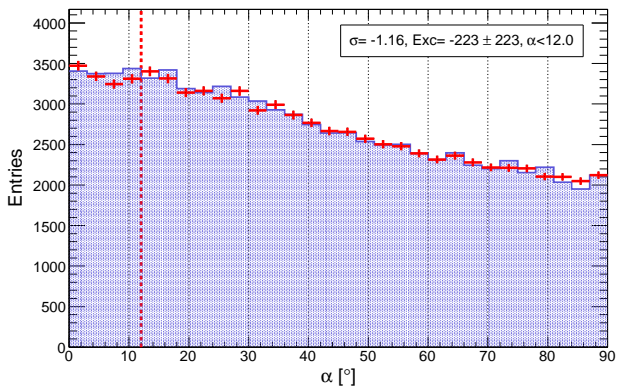}
\caption{Willman~1 $\alpha$-plot as seen by MAGIC in $15.5$~hours above
  a fiducial energy threshold of 100~GeV and using a
  hadronness$<0.15$. The red crosses represent the ON-data sample, the blue
  shaded region is the OFF-data sample normalized to the ON-data
  sample between $30^\circ-80^\circ$. The vertical red dotted line
  represents the fiducial region $\alpha<12^\circ$ where the signal is
  expected. \label{fig:alpha}}  
\end{figure}

No significant $\gamma$-ray excess beyond 100~GeV above the background was
observed in $15.5$ hr of observation of the sky region around
Willman~1, according to Figure~\ref{fig:alpha}, where the
``$\alpha$-plot'' is reported. The $\alpha$-parameter is the angular distance
between the shower image main axis and the line connecting the image
barycenter and the camera center. Due to their
isotropic origin, hadronic events, in case they survive the analysis,
are randomly oriented in the camera both in the ON-data and
OFF-data sample. This is reflected  into a rather smooth
distribution of events in the 
$\alpha$-plot, the nonperfect flatness being due to an increased camera
acceptance for showers with small $\alpha$.
On the other hand, $\gamma$-rays trace back the source, and thus the
orientation  of the shower image is toward the center of the camera.
 Therefore, in case of positive
detection, an  excess of events in the ON-data above the OFF-data
sample is expected at small $\alpha$. A fiducial region
$\alpha<12^\circ$ is chosen where
the signal is assumed with a cut slightly larger than for a pointlike
source to take into account the moderate source
extension. The OFF-data are normalized to
the ON-data in the region where  no signal is expected, 
i.e., between $\alpha=30^\circ$ and $\alpha=80^\circ$.

The significance is calculated using Equation~17 of \citet{Li:1983}. The
number of excess events $N_{exc}(>100$~GeV$)=-223 \pm 
223$ is calculated as the difference between the number of ON-events
and the  number of OFF-events in the fiducial $\alpha-$region. 
The method from \citet{Rolke:2005} was applied to estimate the upper
limit in the number of excess events with a $90$\% confidence level
and including $30$\% of systematic errors, giving as a result
$N^{ul}_{exc}(>100$~GeV$)\sim191.4$ events.  This value is used to
reconstruct the corresponding photon flux for a general $\gamma$-ray
spectrum $S(E)$ using:  
\begin{equation}\label{eq:ul}
\Phi^{u.l.}_{>E_0}=\frac{N^{u.l.}_{exc}}{\int S(E)\,
  A^{cuts}_{eff}(E)\;\textrm{d}E\, 
  \Delta T}\,\int_{E_0} S(E)\textrm{d}E\;,
\end{equation}

\noindent
where $E_0$~GeV is the energy threshold, $A^{cuts}_{eff}(E)$ is the
effective telescope area  and $\Delta T$ is the effective observation
time.  

We applied Equation~\ref{eq:ul} for the four neutralino benchmarks
defined. Results are reported in Table~\ref{tab:results}, where we also
compare $\Phi_{>E_0}^{u.l.}$ with the estimated flux of
Table~\ref{tab:bm}. We also report the upper limits on the \emph{boost
  factors} that are required to match the two fluxes, again
calculated separately for each neutralino model. The boost factor 
is defined as the ratio between the upper 
limit and the theoretical flux, and defines the minimal
boost that the theoretical flux should be subject to in order to allow
for a positive detection of the source. In order to provide results
less dependent of the particular benchmark spectrum, we also
calculated flux upper limits in four different energy bins [100-170,
  170-350, 350-1000, 1000-$20\,000$]~GeV for a generic annihilation
spectrum without cutoff and spectral index $-1.5$. Respectively, the
resulting upper limits are
$[\,9.94,\,4.75,\,0.68,\,0.35\,]$$\times10^{-12}$
photons~cm$^{-2}$~s$^{-1}$.

\begin{table*}[htb!]
\begin{center}
\caption{\label{tab:results}Comparison of estimated integral flux
  above 100~GeV using Equation~\ref{eq:flux} for the benchmarks models
  defined in Table~\ref{tab:bm} and the upper limit in the integral
  flux $\Phi^{u.l.}$ above 100~GeV coming from MAGIC data in units of 
  photons cm$^{-2}$ s$^{-1}$. On the rightmost column, the
  corresponding upper limit on the boost factor $B^{u.l.}$ required to
  match the two fluxes is calculated.}
\vspace{2mm}
\begin{tabular}{cccc}
\tableline\tableline
  BM & $\Phi^{model}(>100\;\mbox{GeV})$ &
  $\Phi^{u.l.}(>100\;\mbox{GeV})$ & $B^{u.l.}$ \\
\tableline
$I'$  & 2.64$\times10^{-16}$ & $9.87\times10^{-12}$ & $3.7\times10^{4}$ \\
$J'$  & 4.29$\times10^{-17}$ & $5.69\times10^{-12}$ & $1.3\times10^{5}$\\
$K'$  & 2.32$\times10^{-15}$ & $6.83\times10^{-12}$ & $2.9\times10^{3}$\\
$F^*$ & 2.09$\times10^{-16}$ & $7.13\times10^{-12}$ & $3.4\times10^{4}$\\
\tableline\tableline
\end{tabular}
\end{center}
\end{table*}

Table~\ref{tab:results} reveals that although we derived upper limits of
the same order of ma\-gni\-tu\-de for the four models 
considered, there are evident differences in the prospects of
detection for different neutralinos. 
The boost factor largely depends on the benchmarks, but the main
differences are connected to the particle physics factor, which varies by
orders of magnitude among the different benchmarks, as shown in
Table~\ref{tab:bm}. The best prospects are for neutralinos in the
$funnel$ region (model $K'$) of the parameter space, for which the
mass is large enough to place the 
cutoff well within the MAGIC energy threshold but still small enough
not to reduce the particle physics factor $\Phi^{PP}$ of
Equation~\ref{eq:phipp} too much. Next, with similar boost requirements
follow the $I'$ and $F^*$ models. In the former case, the effect of the
IB plays an important role at energies close to the cutoff even if the
neutralino mass is 
very close to the MAGIC energy threshold. In the latter case, although
the IB effect is negligible, the signal is very extended in the
energy region suitable for MAGIC, whereas the large mass makes the
flux suppression too large. The worst case scenario is
the co-annihilation neutralino. In this case, even if the IB contribution
is large, the intrinsic total cross section is very low, which makes
the flux very low compared with the others. The IB effect cannot
counteract this intrinsic deficit. 

Although the results of Table~\ref{tab:results} 
are at first glance not looking overly promising
for indirect detection of the discussed models in the immediate future,
one should keep in mind that the estimation of the flux is model dependent
and there are intrinsic effects that could boost up the flux for up
to 1 order of magnitude if the effect of substructures,
astrophysical uncertainties, and/or the inclusion of baryons is 
considered (as already discussed in
Section~\ref{subsec:phipp}) and for several orders of magnitude, if
the DM particle velocity is small enough to give rise to the
Sommerfeld effect.  This could change completely the outlook, as also discussed in \citet{Pieri:2009}.

\section{Conclusion}
\label{sec:conclusions}
In the context of DM searches, we have observed the Willman~1 dwarf 
galaxy with the MAGIC telescope for a total of 15.5 hr  between
March and May 2008. Willman 1 represents one of the best DM-dominated 
systems known in the universe to search for DM at present, according 
to its inferred dynamical properties and distance. No $\gamma$-ray signal 
was detected above an energy threshold of 100~GeV. We 
have obtained different flux upper limits of the order of
$10^{-12}$~ph cm$^{-2}$ s$^{-1}$ separately for four benchmark models 
considered in the framework of mSUGRA. 

Using the latest estimations of its structural parameters to build
the DM density profile as well as the inclusion of the recently
proposed IB mechanism, we calculated the boost factors needed to match
the expected flux values from DM annihilation in Willman~1 with the
upper limits obtained from the data. We can see that boosts in flux in the
order of 10$^3$ are required in the most optimistic scenario
considered. However, uncertainties in the DM distribution, the role of
DM substructure or/and the contribution from the Sommerfeld
enhancement may
reduce this required boost significantly. It is expected that deeper
observations of the Willman~1 dSph with the upcoming
MAGIC~II telescope will allow us to improve the flux limits
presented here by a factor $2-10$. 

\acknowledgments
We would like to thank the Instituto de Astrofisica de Canarias for
the excellent working conditions at the Observatorio del Roque de los
Muchachos in La Palma. The support of the German BMBF and MPG, the
Italian INFN and Spanish MCINN is gratefully acknowledged. This work
was also supported by ETH Research Grant TH 34/043, by the Polish
MNiSzW Grant N N203 390834, and by the YIP of the Helmholtz
Gemeinschaft. Finally, we thanks the anonymous referees for useful
comments. 

{\it Facilities:} \facility{MAGIC}.

\end{document}